\documentstyle[11pt,newpasp,twoside,epsf]{article}
\begin{document}
\title{The detection of spiral arm modulation in the mass
       distribution of an optically
       flocculent galaxy}
\author{Iv\^anio Puerari}
\affil{Instituto Nacional de Astrof\'\i sica, Optica y Electr\'onica,
       Mexico} 
\author{David L. Block}
\affil{Dept of Computational and Applied Mathematics, University
       Witwatersrand, South Africa}
\author{Bruce G. Elmegreen}
\affil{IBM Research Division, T.J. Watson Research Centre, USA}
\author{Jay A. Frogel \& Paul B. Eskridge}
\affil{Dept of Astronomy, The Ohio State University, USA}

\begin{abstract} Spiral arm modulation in NGC 4062 (an optical
       flocculent) is detected for the first time; 
the Fourier spectra of NGC 4062 and
       NGC 5248 (a classic grand design in optical images) are almost identical.
\end{abstract}

\vskip20pt
\noindent Within the framework of the Density Wave Theory (Lin and Shu
1964) and the the modal theory of galactic spiral structure
(Bertin et al. 1989) the
{\it characteristic signature} of a two-armed grand design is the variation
of amplitude with radial distance from the center, due to the interference
of wave packets or modes
which propagate inward and outward, being reflected off a central bulge. 

Symmetric spiral arm amplitude modulations indicative of underlying wave
modes have hitherto only been detected in the grand design galaxies
M51, M81, M100 (Elmegreen et al. 1989) and in the multiple arm galaxy M101
(Elmegreen 1995). In all the cases, blue images (revealing the young
Population I disk) were analysed. On the other hand, near-infrared images
reveal the old stellar Population II disk component of spiral galaxies.
The young Population I disk component may only
constitute 5 percent of the dynamical mass of the disk of a galaxy.
For studying mass distributions of disk galaxies,
near-infrared images are essential (Block and Wainscoat 1991;
Block and Puerari 1999).

We have applied a morphological method, based on the
bidimensional Fourier transform\footnote{The 
bidimensional Fourier method has
been extensively discussed in a number of papers, and the reader
is referred to eg. Consid\`ere and Athanassoula (1982),
Puerari and Dottori (1992), Puerari (1993), amongst others.},
to detect the existence of structures
with a different winding sense (trailing and leading patterns) in the same
galaxy. The Fourier method was applied to the
near-infrared images of two
galaxies which optically could not be more different: one is
flocculent (NGC 4062) whereas the other (NGC 5248) is grand
design. NGC 4062 
displays a duality in spiral structure (see Block et al., this
volume);
its gaseous and stellar disks fully decouple.

In Figure 1, we show the bidimensional Fourier spectra for the bisymmetric
m=2 component of both galaxies.
Note the remarkable similarity between the spectra of both galaxies,
despite the very different optical appearence of these
galaxies (NGC4062 - optically flocculent; NGC5248: optically grand design).
The consequence of two spiral patterns with
{\it different winding sense} in the m=2 component directly implies
spiral arm modulation.

\begin{figure}
\plottwo{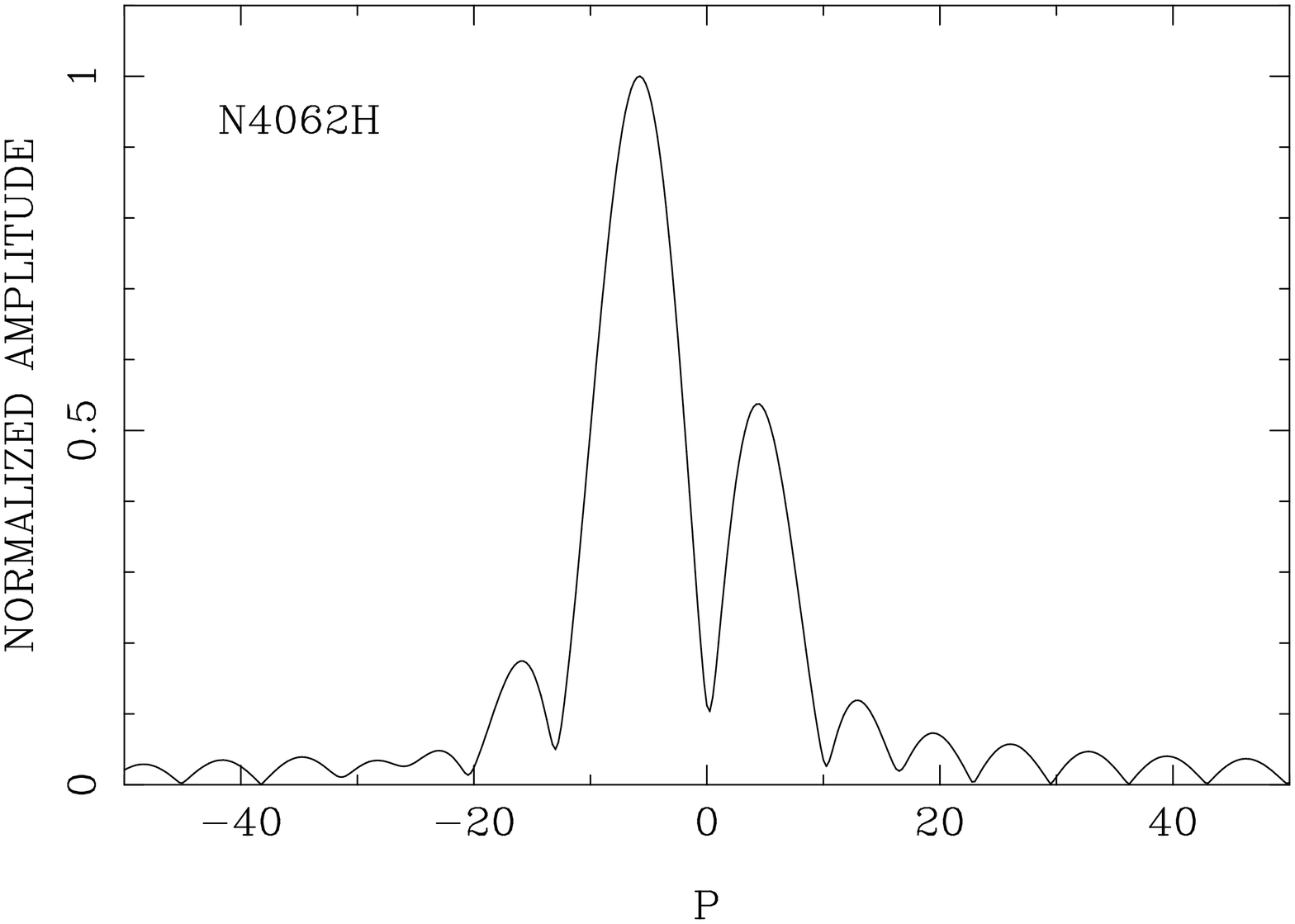}{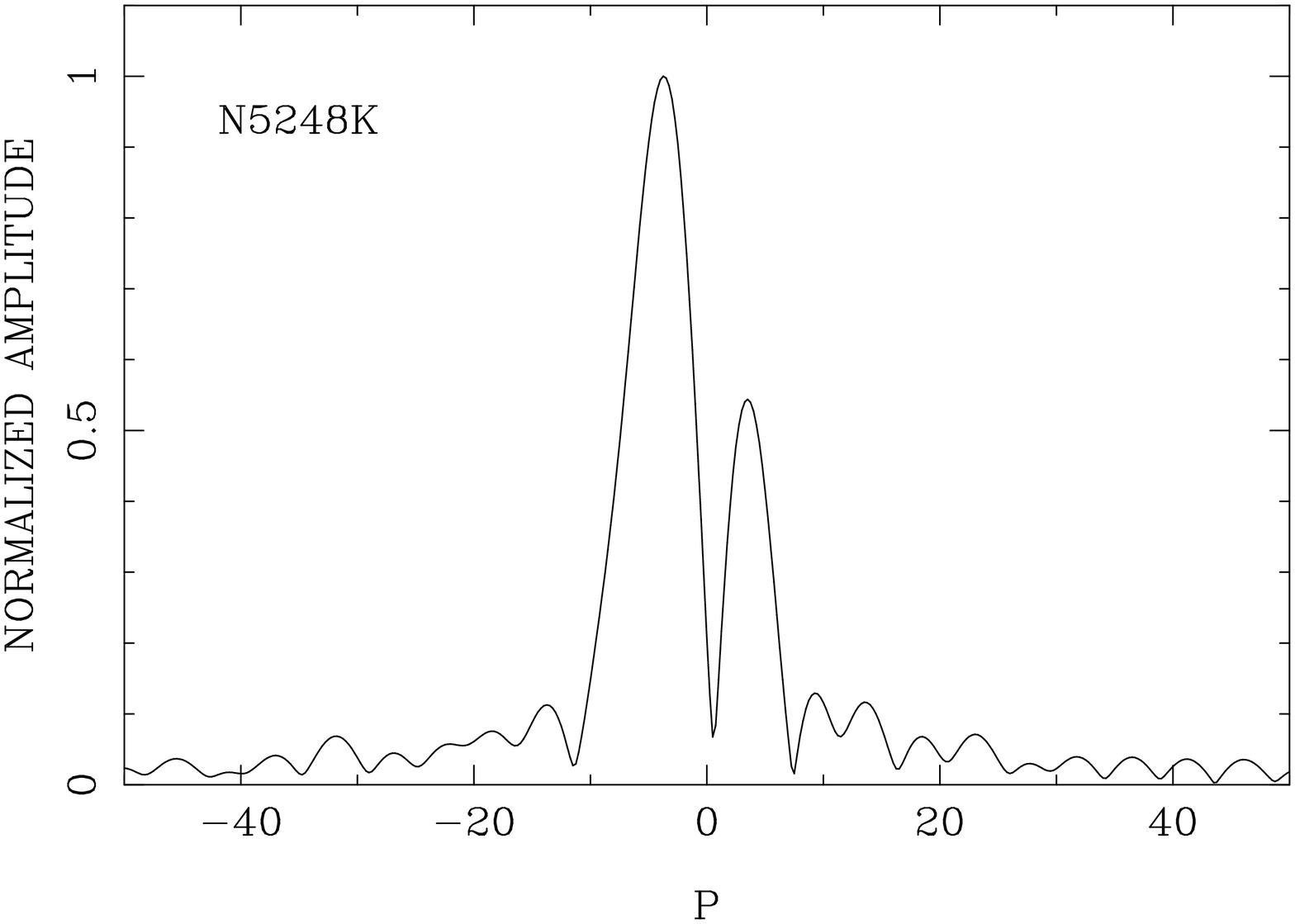}
\caption{An optically flocculent (NGC 4062) and an optically
  `extreme' grand design galaxy (NGC 5248) yield almost identical
  Fourier spectra in the near-infrared regime. Their mass distributions
  must be very similar, despite radically different optical
  morphologies. The m=2 Fourier component is illustrated in both
  cases.}
\end{figure}

A complete and detailed analysis of these galaxies was published
in Puerari et al. 2000 (see also astro-ph/0005345).


\begin{references}
\reference Bertin, G., Lin, C.C., Lowe, S.A., Thurstans, R.P.,
1989, ApJ 338, 78
\reference Block, D.L., Puerari, I., 1999, A\&A 342, 627
\reference Block, D.L., Wainscoat, R.J., 1991, Nature 353, 48
\reference Consid\`ere, S., Athanassoula, E., 1982, A\&A 111, 28
\reference Elmegreen, B.G., 1995, in Physics of the Interstellar Medium and
Intergalactic Medium, eds. A. Ferrara, C.F. McKee, C. Helles and
P.R. Shapiro, ASP Conference Series, 80, 218
\reference Elmegreen, B.G., Elmegreen, D.M., Seiden, P.E., 1989, ApJ 343, 602
%\reference Kalnajs, A.J., 1975, in La Dynamique des Galaxies
%Spirales, ed. L.  Weliachew (Paris Editions du CNRS), 103
%\reference Lau, Y.Y., Lin, C.C., Mark, J.W.-K., 1976, Proc. Natl. Acad. Sci.
%USA 73, 1379
%\reference Lin, C.C., 1970, in The Spiral Structure of Our Galaxy, eds.
%W. Becker and G. Contopoulos (Dordrecht Reidel), 377
\reference Lin, C.C., Shu, F.H., 1964, ApJ 140, 646
%\reference Mark, J.W.-K., 1977, ApJ 212, 645
\reference Puerari, I., 1993, PASP 105, 1290
\reference Puerari, I., Dottori, H.A., 1992, A\&AS 93, 469
\reference Puerari, I., Block, D.L., Elmegreen, B.G., Frogel, J.A.,
Eskridge, P.B., 2000, \aap, 359, 932
\end{references}
\end{document}